\documentclass[aps,prl,twocolumn,groupedaddress]{revtex4}
\usepackage[draft]{hyperref}
\usepackage{amsmath,amssymb}
\usepackage{graphicx}
\usepackage{color}

\begin{document}

\title{Extreme Events in Resonant Radiation from Three-dimensional Light Bullets.}

\author{T. Roger$^{1}$, D. Majus$^{2}$, G. Tamosauskas$^{2}$, P. Panagiotopoulos$^{3}$, M. Kolesik$^{3,4}$, G. Genty$^{5}$, A. Dubietis$^{2}$, D.~Faccio$^{1,*}$}

\address{
$^{1}$School of Engineering and Physical Sciences, SUPA, Heriot-Watt University, Edinburgh EH14 4AS, UK\\
$^{2}$Department of Quantum Electronics, Vilnius University, Sauletekio Ave. 9, Building 3, LT-10222 Vilnius, Lithuania\\
$^{3}$College of Optical Sciences, University of Arizona, Tucson AZ 85721, USA\\
$^{4}$Department of Physics, Constantine the Philosopher University, Nitra, Slovakia\\
$^{5}$Tampere University of Technology, Institute of Physics, Optics Laboratory, FIN-33101 Tampere, Finland
}
\email{d.faccio@hw.ac.uk}

\begin{abstract}
We report measurements that show extreme events in the statistics of resonant radiation emitted from spatiotemporal light bullets. We trace the origin of these extreme events back to instabilities leading to steep gradients in the temporal profile of the intense light bullet that occur during the initial collapse dynamics. Numerical simulations reproduce the extreme valued statistics of the resonant radiation which are found to be intrinsically linked to the simultaneous occurrence of both temporal and spatial self-focusing dynamics. Small fluctuations in both the input energy {\emph{and}} in the spatial phase curvature explain the observed extreme behaviour. 
\end{abstract}


\maketitle
The first direct measurement of extreme or rogue wave in the ocean \cite{draupner} has triggered a large number of studies and renewed interest in what used to be believed as a ``sailor's tale''.  Extreme events have since been observed in many other systems illustrating the wide range of underlying dynamics that can lead to their appearance \cite{rogues1,rogues2}.  In an optical context, the recent observation that solitons with extreme spectral red-shifts can emerge from an incoherent fibre supercontinuum \cite{solli} has attracted considerable attention and subsequent studies have highlighted the role of nonlinear noise amplification in the spontaneous emergence of highly localised structures \cite{breather,peregrine} and the role of collisions in observing long tail statistics \cite{akhmediev,erkintalo,millot}.

While most rogue waves studies in optics have focussed  on the 1+1D fibre case (one spatial dimension and one temporal dimension), the emergence of extreme events during full 3D propagation of light pulses with nonlinear spatio-temporal dynamics has been much less studied \cite{kasp,dub1,steinmeyer,dub3}.  In the normal group velocity dispersion (GVD) regime, the combined action of spatial and temporal self-focusing leads to the formation of a dynamically stable filament characterised by an intense, localised peak surrounded by a weaker reservoir that continuously refuels the central core region \cite{couairon}.  In this regime, long tail statistics in the fluctuations at the spectral edge of a single filament similar to that of an incoherent fibre supercontinuum have been observed \cite{kasp,dub1,dub3}.  More recently, extreme events in the multi-filament regime where localized structures emerge from filaments merging have been measured experimentally\cite{steinmeyer}. 

Strictly speaking, 3D stable solitons do not exist and the physical structure closest to a soliton is the 2D Townes spatial profile \cite{townes,malomed} which is unstable and leads to self-similar collapse of the laser beam in the normal GVD regime when the beam power exceeds the critical power for self-focusing \cite{gaeta}. On the other hand, in the anomalous GVD all three dimensions contribute to the spatio-temporal collapse with formation of light bullets \cite{bullets}.  Although such light bullets lend themselves to a tempting analogy with 1D fundamental solitons, recent work has shown that in fact 3D light bullets correspond to a form of polychromatic (weakly localised) Bessel beam that emerges spontaneously during the collapse phase of an initially Gaussian-shaped wavepacket \cite{arxiv}.  Despite this difference, light bullets do exhibit remarkable similarities with 1D solitons: (i) they appear to propagate quasi-undistorted without pulse splitting as observed in the normal GVD  regime; (ii) temporal compression may occur in a fashion similar to soliton compression and (iii) their propagation in the presence of higher-order dispersion perturbation is accompanied by the emission of a resonant radiation (RR) often referred to as a dispersive wave \cite{bullets,bullets2,RR}.

In this letter, we report on the observation of optical rogue waves associated with the emission of extreme RR during the formation of 3D light bullets in a nonlinear crystal induced by the spatio-temporal coupling of fluctuations inherently present on the input beam.  The deterministic spatio-temporal dynamics are central to the rogue characteristics of the RR emission as a result of the steep shock front that forms on the trailing edge of the pulse during the initial collapse phase and whose gradient is highly sensitive to fluctuations in both the input energy and spatial phase curvature.  Our experimental results are confirmed by full 3D numerical simulations.  We anticipate that similar mechanisms and extreme events can emerge in other systems where temporal and spatial dynamics are coupled through a nonlinear collapse event. 

The emission of resonant radiation was originally described in the 1+1D context as a form of soliton instability that occurs in the presence of higher order dispersion  \cite{RR} and more recently interpreted in terms of cascaded four-wave mixing dynamics \cite{erkintalo}.  The frequency $\omega$ of the RR is determined by a dispersion relation $k(\omega)=k(\omega_0)+(\omega-\omega_0)/v$ where $k$ represent the wave-vector, and $\omega_0$ and $v$ are the central frequency and group velocity of the soliton, respectively \cite{RR}.  The spectral signature of the RR is typically observed as a blue shifted emission in the normal GVD range.  Similar RR emission predicted by the very same phase-matching relation, occurs in 3D both for pulses in the anomalous and normal GVD regimes \cite{born_miro2,born_rubino}.  For normal GVD pumping however, significant continuous spectral broadening may occur simultaneously and it is only for a pump in the anomalous GVD  that the RR appears as an isolated spectral peak with strong similarities to the 1D soliton perturbation as observed in a variety of bulk media \cite{bullets2,jens,dub2}. 

Light bullets were generated by focussing into an 8 mm long sapphire crystal with a $f=+100$ mm lens 100 fs tunable pulses from an optical parametric amplifier (TOPAS-C, Light Conversion Ltd.) operating at 1 kHz repetition rate and for an input energy of 2.5 $\mu$J in agreement with previous observations \cite{arxiv}.  The input energy was monitored with a pyroelectric detector. The energy and spectrum of the RR at the crystal output on the other hand were measured with a calibrated silicon photodiode and a fibre spectrometer (QE65000, Ocean Optics), respectively.  The detection system was synchronized with the laser pulse allowing for single-shot characterisation.

\begin{figure}[t]
\centering
\includegraphics[width=8.5cm]{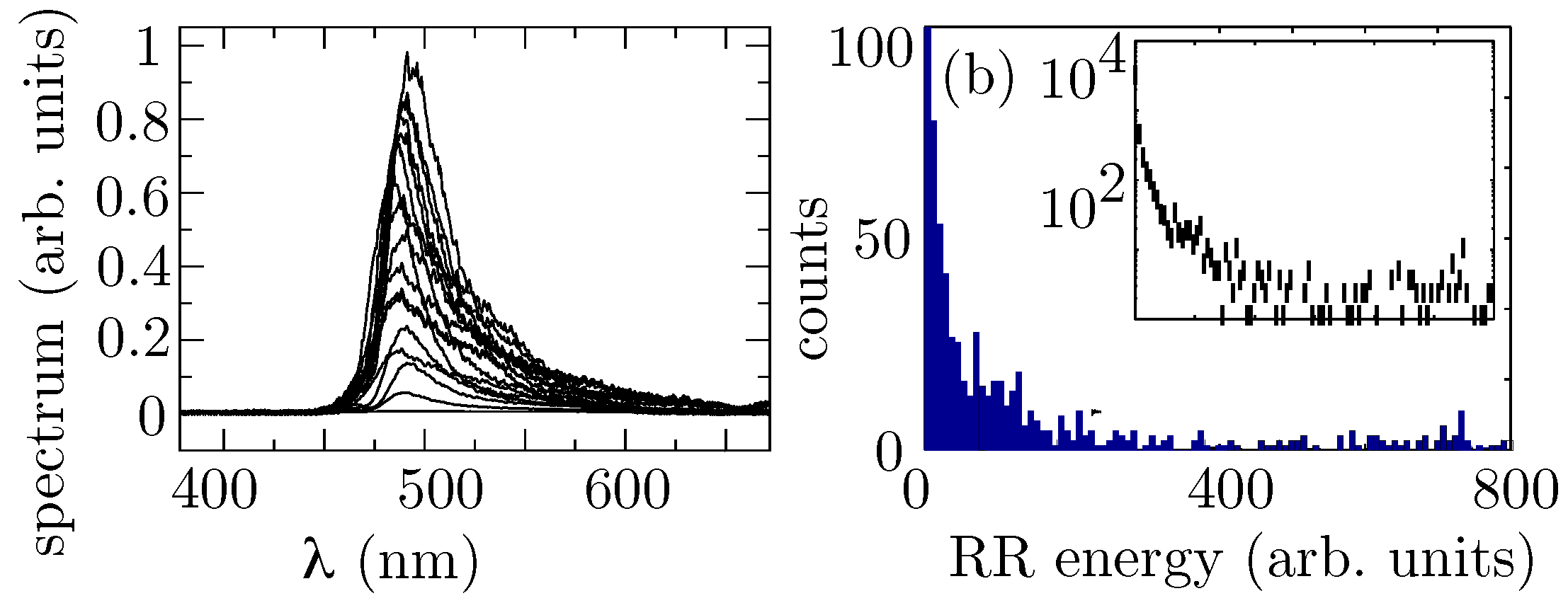}
\caption{(a) Selection of 20 representative spectra recorded at nominally constant input energy and focussing conditions. (b) Statistical distribution of the RR energy (calculated from the integrated spectrum) and zoomed in along the vertical axis on the lower count rate range (0-100) - the full vertical range of data is shown in log scale in the inset.}
\label{exp1}
\end{figure}

The light bullets produce a relatively intense and isolated resonant radiation in the visible region of the spectrum and which shows a clear flickering and unstable behaviour with occasional events of very high intensity.  Selected single-shot RR spectra measured for an input wavelength of 2.1 $\mu$m and shown in Fig.~\ref{exp1}(a) illustrate how the RR emission wavelength is relatively stable but that there are indeed significant intensity fluctuations.  We emphasise that optical parametric amplifier operates in a relatively stable regime with a 1.02\% standard deviation of the energy fluctuations.  Full statistical analysis of the RR energy performed by integrating the individual RR spectra of 3000 laser shots reveals a highly skewed distribution as illustrated in Fig.~\ref{exp1}(b) (the inset shows the full data in log scale). Defining the significant energy as the mean of the highest one third energies recorded \cite{erkintalo2,zaviyalov} (directly adapted from the significant wave height criterion traditionally used in hydrodynamics \cite{height}), events whose energy exceeds by a factor of 2 $W_s$ can be considered as rogue \cite{rogue_def}.  In our measurements, we observe extreme fluctuations that exceed the significant energy $W_s$ by more than a factor 5, a clear indication of the presence of rogue events.  At first sight such extreme statistics in the RR intensity may seem surprising as the propagation dynamics of the RR wave are typically linear due to the strong dispersion at the RR wavelength which results in the rapid walk-off from the main pump pulse and thus background-free propagation. 

\begin{figure}[t]
\centering
\includegraphics[width=6cm]{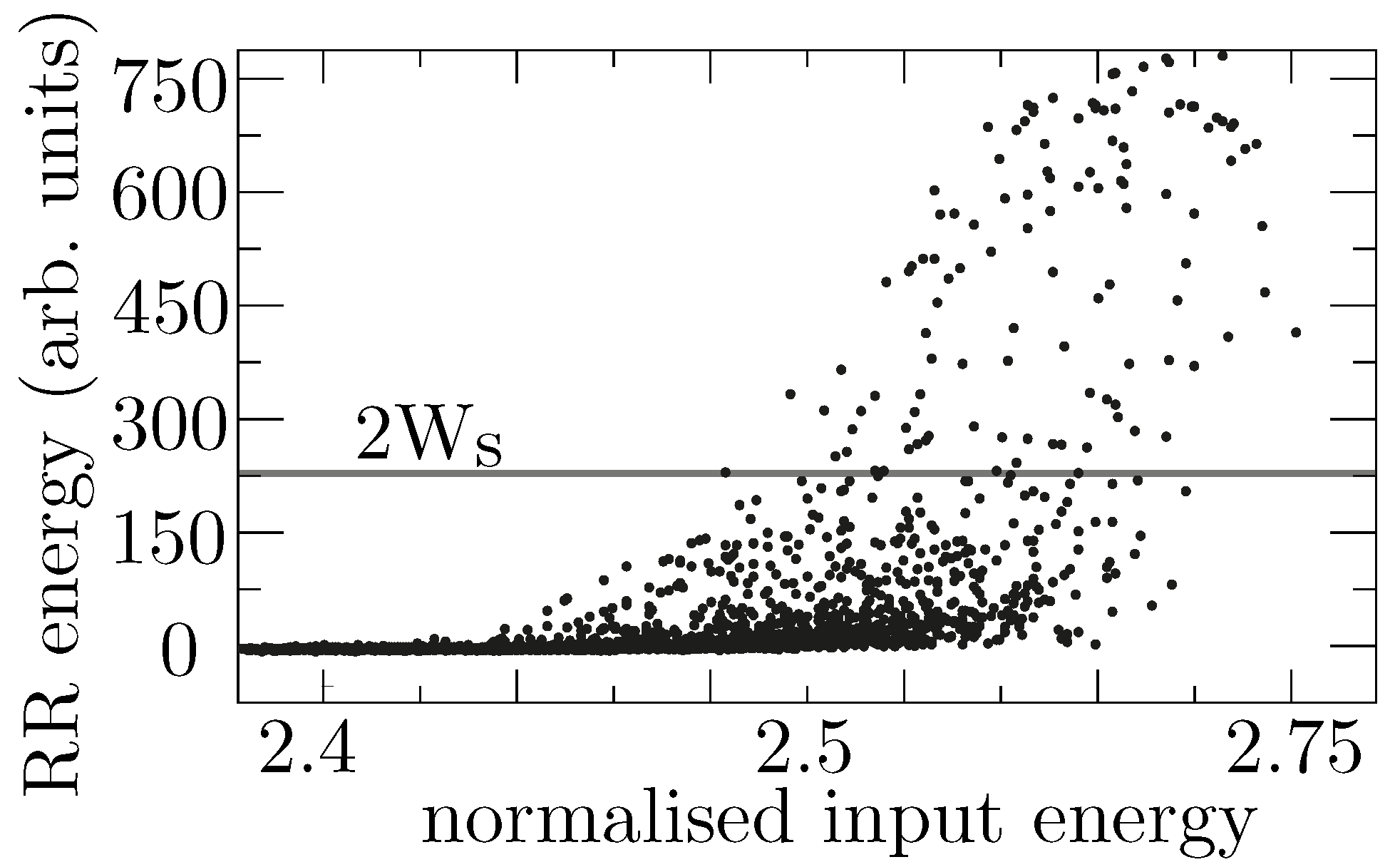}
\caption{Experimentally recorded RR energy as a function of input energy. The grey line indicates 2x the significant energy mark $W_s$.}
\label{exp2}
\end{figure}

\begin{figure}[t]
\centering
\includegraphics[width=8cm]{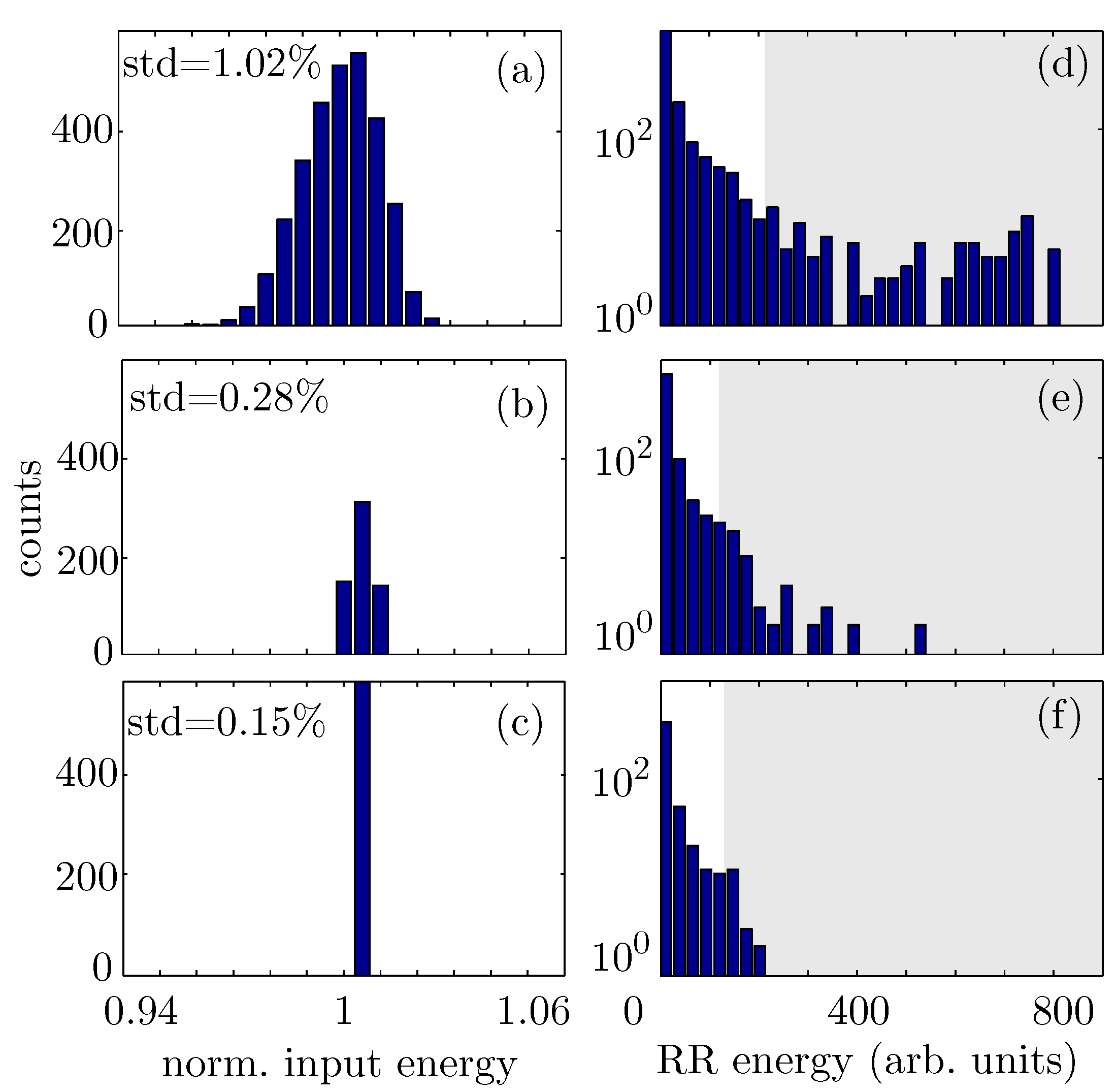}
\caption{(a) Experimentally measured energy distribution of the input pump pulse (normalised to its mean value). (b) and (c) The same data, post-selected so as to keep only data with a smaller fluctuation (standard deviations of 0.28\% and 0.15\%, respectively) around a fixed mean value. (d)-(f) Histograms of the RR energy counts for the input energy distributions shown in (a)-(c): selecting input pump pulses with a narrow energy distribution clearly also narrows the RR energy L-shaped distribution. The shaded areas indicate events that are larger than $2W_s$. }
\label{exp_hist}
\end{figure}

In order to clarify the origin of the extreme fluctuations observed in the RR energy, we plot in Fig.~\ref{exp2} the change in the RR energy as a function of the input pulse energy.  The grey horizontal line marks the 2$W_s$ limit and there are clearly many rogue events that lie above this line.  The fact that the number of rogue events increases with the input energy fluctuations suggests that the RR energy is related to the input energy via a steep nonlinear transfer function \cite{dub4} that converts the narrow (Gaussian) distribution of the input pulse energy into a distinct long tailed distribution of the RR energy.  To further confirm this hypothesis, we plot in in Fig.~\ref{exp_hist} the histogram of the measured RR energy for decreasing input energy fluctuations (using data post-selection) and we observe a gradual reduction in the number of events that populate the tail of statistical distribution and suggesting deterministic dynamics for the occurrence of the rogue RR events.  Significantly, for a fixed input energy, we still obtain a skewed (L-shaped) distribution of the RR energy, as can be seen from the dots taken along a single vertical line in the RR energy scatter of Fig.~\ref{exp2}.  This means that energy fluctuations alone can not explain our observations and that there must be at least one other relevant parameter that plays a role.  These features are confirmed over the wider 1.7-2.2 $\mu$m wavelength range for pump pulses were observed for input wavelengths where RR with similar energy characteristics to those shown in Figs.~\ref{exp1} and \ref{exp2} were observed.
 
The amplitude of the RR emitted by 3D light bullets is directly proportional to the steepness of the shock front that forms on the trailing edge of the collapsing input pulse as it converges towards the formation of a stable light bullet \cite{born_miro,born_rubino}.  It is then natural to link the extreme fluctuations in the RR energy to a similar rogue behaviour in the temporal gradient of the light bullet.  In order to confirm the physical origin of the rogue statistics and gain a deeper understanding of the underlying dynamics, we have performed a series of numerical simulations based on the unidirectional pulse propagation equation (UPPE) \cite{uppe} using the same input parameters as in the experiments.  A defining feature of our simulations is that the rogue statistics are not seeded by quantum noise.  This can be justified on the basis that there do not appear  to be any nonlinear mechanism such as four-wave-mixing or modulation instability that would be sensitive to such fluctuations but instead the RR emission is a coherent process resulting from the pump pulse itself \cite{born_miro2,born_rubino}.  We therefore assume that the light bullet and associated RR emission is fully deterministic \cite{det1,det2,dub4} so that the experimentally observed statistical behaviour and rogue events may be reproduced by accounting for small fluctuations or changes in the input parameters which effectively leads to a stochastic system requiring statistical treatment.  Most importantly, we allow for variations both in the input energy and in the profile of the spatial phase curvature.  The extent of the spatial phase variation in the incident plane of the nonlinear crystal considered in the simulations is very weak, of the order of tens of meters at the laser output (i.e. before the actual experimental setup).  Such small variations can be easily explained e.g. from air turbulence along the beam path and from fluctuations from the OPA itself due to the nonlinear processes involved in the OPA operation (supercontinuum generation and optical parametric amplification). 

\begin{figure}[t]
\centering
\includegraphics[width=8.5cm]{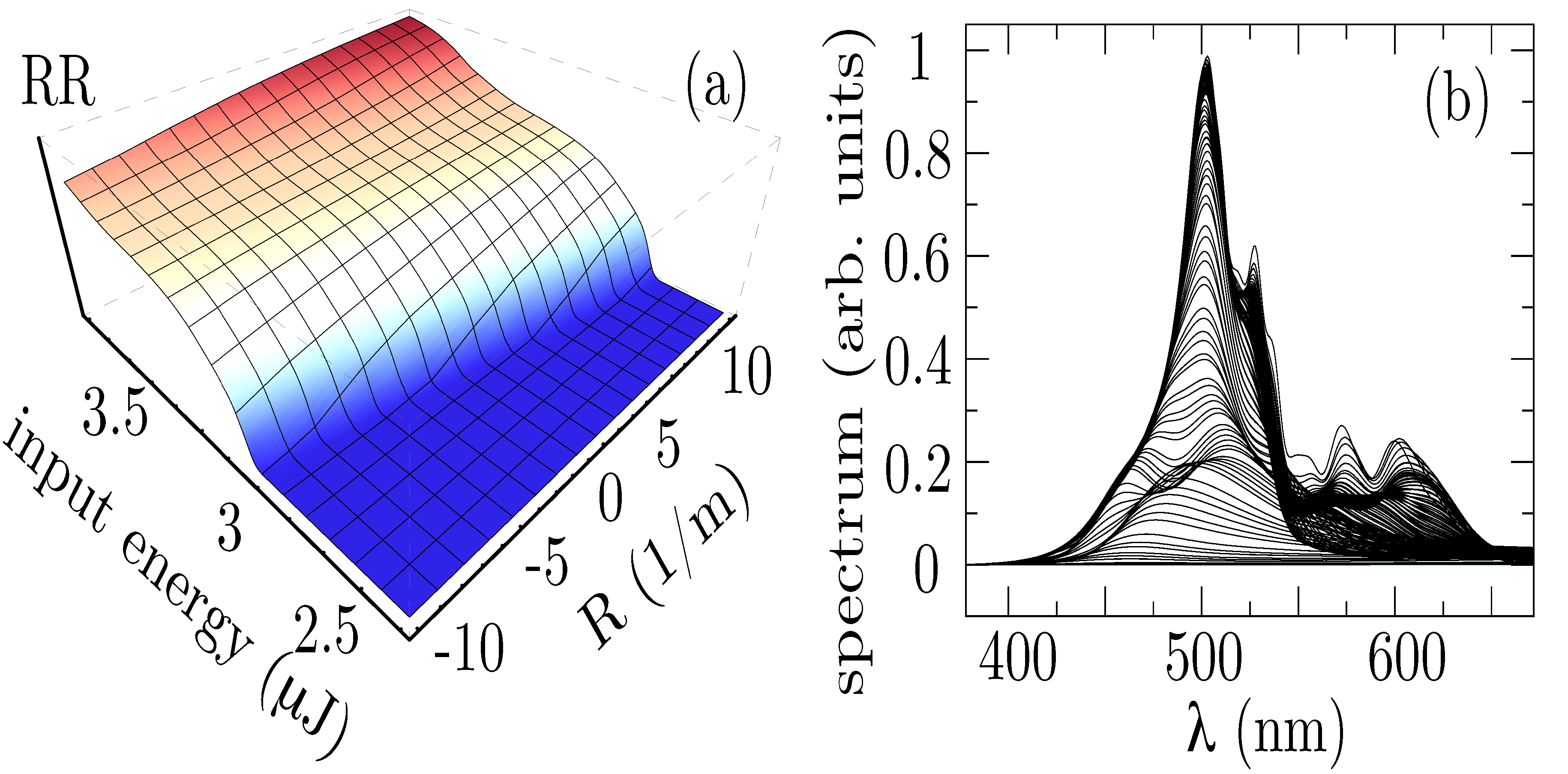}
\caption{(a) Numerically simulated RR energy vs. input energy and spatial phase curvature. (b) Superimposed RR spectra corresponding to all input energy and spatial phase curvature values used shown in (a).}
\label{num1}
\end{figure}

\begin{figure}[t]
\centering
\includegraphics[width=8.5cm]{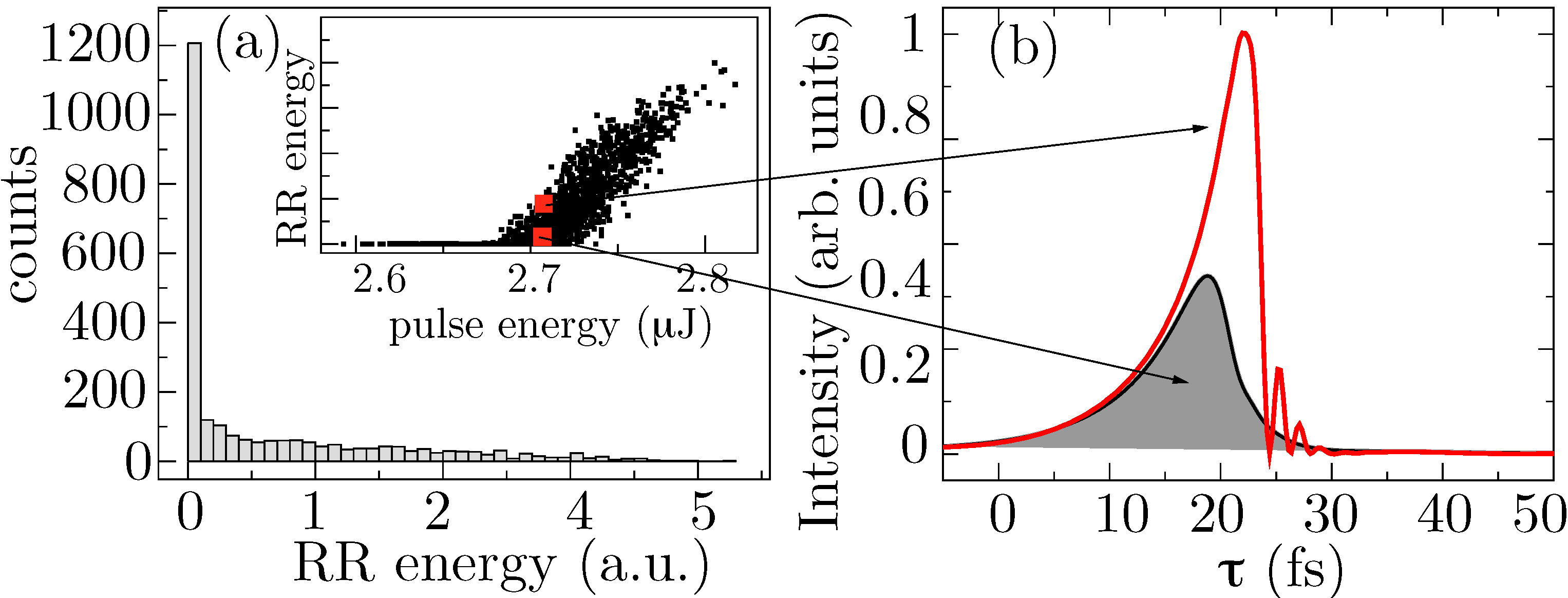}
\caption{Numerical results: (a) statistics of the RR energy (inset shows RR energy versus input energy for zero-phase curvature). (b) On-axis temporal profiles of the light bullet corresponding a fixed input energy but different phase curvature [red squares in the inset of (a)] and giving rise to RR with low energy and high energy.}
\label{num2}
\end{figure}

The simulated RR energy is clearly a nonlinear function of the laser pulse energy and spatial phase curvature as shown in Fig.~\ref{num1}(a).  Significantly, when we plot the RR spectra corresponding to each point of the 2D surface plot of Fig.~\ref{num1}(a) we find a striking resemblance between the simulated spectra and those observed experimentally [compare Fig.~\ref{num1}(b) with Fig.~\ref{exp1}(a)] and see very large fluctuations in the RR energy.  The histogram in Fig.~\ref{num2}(a) showing the statistics of the RR energy computed over the full parameter space  of Fig.~\ref{num1}(a), assuming normal distributions of pulse energy and curvature fluctuations, is also in excellent agreement with the experimental results in Fig.~\ref{exp1}(b).  The projection of the RR energy on the plane of zero-phase curvature that gives the RR energy vs. input pulse energy [see inset in Fig.~\ref{num2}(a)] shows reduced variations compared to the experimentally observed fluctuations of Fig.~\ref{exp2}, which confirms the role of the initial spatial phase curvature fluctuations in observing rogue statistics.  

The importance of the nonlinear coupling between the spatial and temporal dynamics in the propagation of light bullets leading to the observation of rogue statistics is highlighted in Fig.~\ref{num2}  where we show the temporal profile of the simulated light bullet at the crystal output for a fixed input energy of 2.7 $\mu$J in the case of a low and large value of the RR energy.  We can see how the development of a steeper shock front on the trailing edge of the light bullet gives rise to a RR with a much higher energy and that in this case the discrepancy in the shock from gradient is caused by an initial variation in the spatial phase curvature.  Yet, it is important to realise that it is \textit{only} by including variations in the \textit{energy} and the \textit{spatial phase curvature} of the input beam that we are able to reproduce numerically the experimental results so that variations in the spatial phase curvature alone can not explain the experimental behaviour.  This is because the steepness of the shock front is deeply rooted into the spatio-temporal coupling such that initial fluctuations in the input energy and spatial phase curvature leads to significantly larger discrepancies in the gradient of the shock steepness.

In conclusion,  laser pulses propagating in a bulk nonlinear medium with anomalous GVD undergo a spontaneous reshaping into a 3D light bullet that, although being a distinctly non-solitonic pulse \cite{arxiv}, exhibits characteristics features such as the emission of resonant radiation that are commonly observed for 1D solitons.  Yet, the additional spatial degree of freedom leads to significantly richer dynamics and, in particular, to a spatio-temporal collapse that is characterised by rogue wave statistics in the amplitude of the resonant radiation.  These statistics originate from the RR generation process itself, i.e. from the formation of a temporal shock front on the trailing edge of the pulse which is shown to exhibit extreme sensitivity to the input conditions.  

We have generalised an approach put forward in~\cite{dub4} yielding simple, yet effective, method to numerically model the rogue statistics.  Importantly, it provides immediate physical insight by identifying a system-specific transfer function that relates the observed dynamics to the relevant physical parameters that are underpinning the rogue-event generation process.  We believe that other classes of deterministic systems where extreme value statistics are observed can be described in an analogous way, and that the functional form and especially the dimensionality of their corresponding transfer functions can be used to classify different types of rogue events.  We anticipate that the interaction of multiple light bullets that could be spontaneously generated at higher input pulse powers would lead to similar collision processes to those suggested to be responsible for rogue waves in other nonlinear systems \cite{erkintalo,millot}. 

D.F. acknowledges financial support from the European Research Council under the European Unionâ Seventh Framework Programme (FP/2007-2013)/ERC GA 306559 and EPSRC (UK, Grant EP/J00443X/1). D.M., G.T. and A.D. acknowledge financial support from the European Social Fund under the Global Grant measure (Grant No. VP1-3.1-\v{S}MM-07-K-03-001).  G.G. acknowledges support from the Academy of Finland (projects 130099 and 132279). 
P.P. and M.K. were supported by USA AFOSR (Grant No. FA9550-10-1-0561).
Simulation software used was developed with funding from AFOSR, FA9550-11-1-0144. 
 

\end{document}